\documentclass[twocolumn,pra]{revtex4}
\usepackage{amssymb}
\usepackage{amsfonts}
\usepackage{amsmath}
\usepackage{graphicx}

\begin{document}

\title{Maximally non-Markovian quantum dynamics without environment backflow
of information}
\author{Adri\'{a}n A. Budini}
\affiliation{Consejo Nacional de Investigaciones Cient\'{\i}ficas y T\'{e}cnicas
(CONICET), Centro At\'{o}mico Bariloche, Avenida E. Bustillo Km 9.5, (8400)
Bariloche, Argentina, and Universidad Tecnol\'{o}gica Nacional (UTN-FRBA),
Fanny Newbery 111, (8400) Bariloche, Argentina}
\date{\today }

\begin{abstract}
The degree of non-Markovianity allows to characterizing quantum evolutions
that depart from a Markovian regime in a similar way as Schmidt number
measures the degree of entanglement of pure states. Maximally non-Markovian
dynamics are the analogous of maximally entangled states [D. Chru\'{s}ci\'{n}%
ski and S. Maniscalco, Phys. Rev. Lett. \textbf{112}, 120404 (2014)]. Here,
we demonstrate that there exists a class of maximally non-Markovian quantum
evolutions where the associated environment (degrees of freedom not
belonging to the system) obeys a Markovian (memoryless) dynamics, which in
turn is unperturbed by the system state or dynamics. These properties imply
the absence of any \textquotedblleft physical environment-to-system backflow
of information.\textquotedblright\ Non-Markovian features (as usual in
quantum systems coupled to dissipative classical degrees of freedom) arise
from a unidirectional dependence of the system dynamics on the reservoir
states. %The underlying quantum-classical dynamics can be read as a
%(collisional model with a finite number of events).
\end{abstract}

\maketitle

%\title{Maximally nonMarkovian quantum evolutions from interaction with classical degrees of freedom}
%\title{Maximally nonMarkovian quantum evolutions from interaction with classical fluctuations}
%\title{Classical stochastic representation of Maximally non-Markovian quantum dynamics}
%\title{Memoryless arranges leading to Maximally non-Markovian quantum dynamics}
%\title{Maximally non-Markovian quantum evolutions from memoryless Markovian chain dynamics}
%\title{Memoryless realization of Maximally non-Markovian quantum dynamics}
%\title{Non-Markovian quantum dynamics without exchange of information}
%\title{Maximally non-Markovian quantum dynamics without system-reservoir backflow of information}
%\title{Maximally non-Markovian quantum dynamics without backflow of information}
%\title{Maximally non-Markovian quantum dynamics from memoryless reservoirs}
%\title{Maximally non-Markovian dynamics induced by Markovian degrees of freedom}
%\title{Maximally non-Markovian quantum dynamics without backflow of information}

%03.65.Yz Decoherence, Open systems,
%03.65.Ta Foundation of quantum mechanics: measurement theory
%42.50.Lc Quantum fluctuations, quantum noise, and quantum jumps
%05.40.Ca Noise
%05.40.-a Fluctuation phenomena, random processes, noise, and Brownian motion
%05.60.Gg Quantum transport

\section{Introduction}

In contrast to classical stochastic systems, for open quantum systems \cite%
{breuerBook} the definition of memory, or equivalently departure from a
memoryless (Markovian) regime \cite{breuerBook,vega}, is much more subtle.
The theory of quantum dynamical semigroups \cite{alicki} gives a rigorous
basis for defining Markovian dynamics, where the time evolution of the
system density matrix is given by the so called Lindblad equations (or
Gorini-Kossakowski-Sudarshan-Lindblad equations). Assuming that all
available information is encoded in the system density matrix, any departure
in the properties of its propagator with respect to that of a quantum
dynamical semigroup may be utilized as an indicator of the presence of
memory, or equivalently, for defining a \textit{quantum non-Markovian regime}
\cite{BreuerReview,plenioReview}.

The previous point of view was introduced in the seminal contribution of
Breuer, Laine, and Piilo \cite{BreuerFirst}, where the memory indicator is
given by the (non-monotonous) behavior of the distinguishability between two
initial states. Since then, many other indicators and measures were
introduced, such as for example based on the divisibility of the propagator 
\cite{cirac,rivas,HouSeparable,operationalDivisibility}, geometry of the set
of accessible states \cite{geometrical}, negativity of the dissipative rates
in a canonical form of the quantum master equation \cite{canonicalCresser},
non-Markovianity degree \cite{DarioSabrina,franca}, spectra of the dynamical
map \cite{spectra}, quantum regression theorem \cite{QRTVacchini}, power
spectrum \cite{piilo}, Fisher information flow \cite{fisher}, mutual
information \cite{mutual}, fidelity \cite{fidelity}, and accessible
information \cite{brasil} just to name a few \cite{BreuerReview,plenioReview}%
. In addition, many aspects were analyzed such as for example the relation
with the definition of non-Markovian classical stochastic processes \cite%
{Classical,Hybrid}, and the correspondence between the different
(inequivalent) non-Markovian indicators and measures \cite%
{dario11,cresser,luo,dario14,dario17,Acin}.

The authors of Ref. \cite{BreuerFirst} also introduced a new perspective for
understanding non-Markovian or memory effects: one can read any departure
from a Markovian regime (measured, in general, with any of the previous
indicators) in terms of a\textit{\ backflow of information} from the
reservoir to the system. Instead, the memoryless (Markovian) case
corresponds to a unidirectional loss of information from the system to the
reservoir. This conceptual frame was criticized in Ref.~\cite{petruccione}.
On the other hand, there are physical situations where it has a clear
meaning. For example, in a Markovian regime, an excitation of a quantum
optical transition is transferred unidirectionally from the system to the
environment \cite{breuerBook}. In the other extreme, for a two-level system
interacting with a quantized harmonic oscillator (Jaynes-Cumming model \cite%
{breuerBook}) an excitation is continuously transferred back and forth
between the (two-level) system and the environment (quantum harmonic
oscillator). Spin-boson models \cite{breuerBook} interpolate between both
extreme behaviors. This kind of bidirectional\textit{\ physical backflow of
information,} which is mediated between the system and the environment by a
physical variable such as energy or heat, was analyzed in Refs. \cite%
{EnergyBackFLow,Energy,HeatBackFLow} and recently studied experimentally in
Ref. \cite{breuerDrift}.

It is important to notice that memory effects (as detected by any of the
proposed non-Markovian indicators based on the system dynamics) may happen
without a \textquotedblleft physical environment-to-system backflow of
information.\textquotedblright\ For example, the environment, defined by 
\textit{all degrees of freedom not belonging to the system,} may has its own
Markovian (time-memoryless) dynamics which in turn is unperturbed by the
system dynamics or state. Hence, even when the system dynamics is
non-Markovian, the environment (its partial dynamics) is completely unaware
of its coupling with the system.

A class of dynamics with the previous property is given by quantum systems
coupled to environments modeled through stochastic Hamiltonians whose
fluctuations are written in terms of classical non-white (Gaussian) noises.
Even when the system may develops strong non-Markovian features \cite%
{GaussianNoise}, the environment (the noise) is unaffected by the system. In
addition, quantum systems coupled to dissipative classical degrees of
freedom \cite{LindbladRate} may lead to the same situation, that is, the
system dynamics is non-Markovian while the environment follows a Markovian
evolution that is unaffected by the system state. For example, in Ref. \cite%
{PostMarkovian} a class of time-convoluted non-Markovian master equations
were derived over the previous basis, where the non-Markovian indicator is
the relative entropy with respect to the stationary state. In addition, a
class of non-Markovian collisional models \cite{collisional} can be
recovered in a similar way, where a quantum ancilla system\ (part of the
environment) in a bipartite embedding follows its own independent quantum
Markovian Lindblad evolution \cite{embedding}. More recently, master
equations leading to eternal non-Markovianity, a Lindblad-like equation with
a time-dependent rate that is negative at all times \cite{canonicalCresser},
were derived on similar grounds \cite{eternal}. The underlying (Markovian)
environment dynamics can be read in alternative ways such as random
dephasing channels and Markov chains \cite{LindbladRate}.

The main goal of this paper is to extend the previous results \cite%
{PostMarkovian,embedding,eternal} to the case of \textit{maximally
non-Markovian quantum evolutions}. This kind of dynamics was introduced by
Chru\'{s}ci\'{n}ski and Maniscalco in Ref.~\cite{DarioSabrina}. Studying the
positivity of the propagator between two arbitrary times in an extended
Hilbert space (divisibility), the authors defined a \textit{non-Markovianity
degree}. This parameter is\ the analog of Schmidt number in entanglement
theory, and allows to compare (to rank) different non-Markovian evolutions.
Maximally non-Markovian quantum evolutions are the analog of maximally
entangled states. Here, we demonstrate that there exists of class of such
kind of extreme non-Markovian dynamics that are induced by environments
whose own dynamics is unperturbed by the system state, that is, without
existing a physical environment-to-system backflow of information. The
evolutions rely on simple quantum-classical hybrid arranges (collisional
models \cite{collisional} with a finite number of events), where the
classical degrees of freedom (the environment) follow their own local in
time Markovian evolution \cite{LindbladRate}. Similarly to the results of
Refs.~\cite{PostMarkovian,embedding,eternal}, non-Markovian effects arise
due to a unidirectional dependence of the system evolution on the
environment states.

The paper is organized as follows. In Sec. II, different examples of
maximally non-Markovian evolutions are introduced. In Sec. III, the examples
are derived from hybrid quantum-classical dynamics. In Sec. IV, the more
general underlying quantum-classical dynamics that lead to non-Markovian
effects without a physical environment backflow of information are
characterized. Sec. V is devoted to the Conclusions. Information that
supports the main obtained results is provided in the Appendixes.

\section{Maximally non-Markovian evolutions}

Maximally non-Markovian evolutions saturate the degree of non-Markovianity.
Explicit examples that satisfy the required conditions found in \cite%
{DarioSabrina} are provided. The underlying formalism is briefly reviewed in
Appendix A. Both dephasing and random unitary maps are given.

\subsection{Dephasing channel}

Consider a qubit system whose density matrix $\rho _{t}$ follows the
evolution%
\begin{equation}
\frac{d\rho _{t}}{dt}=\frac{1}{2}\gamma (t)(\sigma _{z}\rho _{t}\sigma
_{z}-\rho _{t}),  \label{LindbladDispersivo}
\end{equation}%
where $\gamma (t)$ is a time-dependent rate and $\sigma _{z}$ is the $z$%
-Pauli matrix. The solution can be written as%
\begin{equation}
\rho _{t}=\left( 
\begin{array}{cc}
\rho _{11} & \rho _{12}e^{-\Gamma (t)} \\ 
\rho _{21}e^{-\Gamma (t)} & \rho _{22}%
\end{array}%
\right) ,  \label{solution}
\end{equation}%
where $\Gamma (t)\equiv \int_{0}^{t}\gamma (\tau )d\tau .$ Eq. (\ref%
{solution}) shows the dephasing property of the solution map, that is, only
coherences are affected.

The evolution is completely positive if and only if $\Gamma (t)\geq 0.$
Furthermore, the evolution is maximally non-Markovian if \cite{DarioSabrina}%
\begin{equation}
\lim_{t\rightarrow \infty }\Gamma (t)=\lim_{t\rightarrow \infty
}\int_{0}^{t}\gamma (\tau )d\tau =0.  \label{condition}
\end{equation}%
This simple condition implies the extreme non-Markovian property $%
\lim_{t\rightarrow \infty }\rho _{t}=\rho _{0}$ [see Eq. (\ref{solution})].
Hence, \textit{in the long time limit (stationary state) the initial density
matrix is recovered}.

In order to fulfill the previous condition, we propose the following time
dependent rate%
\begin{equation}
\gamma (t)=2\gamma \frac{1-\gamma t}{e^{\gamma t}-2\gamma t},
\label{RateDispersivo}
\end{equation}%
where the free scaling parameter $\gamma >0$ determines the initial rate
value, $\lim_{t\rightarrow 0}\gamma (t)=2\gamma .$ Its time integral is
given by%
\begin{equation}
\Gamma (t)=\ln \left[ \frac{1}{1-2\gamma te^{-\gamma t}}\right] ,\ \ \ \ \ \
\ \ \lim_{t\rightarrow \infty }\Gamma (t)=0,  \label{GamaDispersivo}
\end{equation}%
which is positive, $\Gamma (t)\geq 0,$ and also satisfies condition~(\ref%
{condition}). Hence, the corresponding evolution\ [Eq. (\ref%
{LindbladDispersivo})] is completely positive and maximally non-Markovian.
In Fig.~1 we plot both the rate $\gamma (t)$ and its time integral $\Gamma
(t).$ The time dependent rate is a smooth function that does not present any
divergence and consistently assumes positive and negative values.%
%figura1%figura%figura%figura%figurav%figura%figura%figura%figura%figura%figura%figura%figura%figura%figurav%figura%figura%figura%figura%figura
%figura%figura%figura%figura%figurav%figura%figura%figura%figura%figura%figura%figura%figura%figura%figurav%figura%figura%figura%figura%figura
\begin{figure}[tbp]
\includegraphics[bb=13 56 740 600,angle=0,width=7.cm]{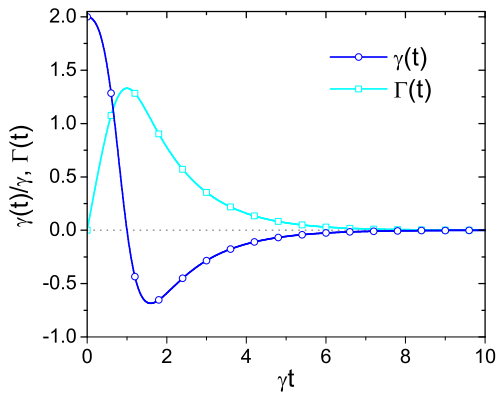}
\caption{Time dependent rate $\protect\gamma (t)$ and its integral $\Gamma
(t),$ Eqs.~(\protect\ref{RateDispersivo}) and (\protect\ref{GamaDispersivo})
respectively, as a function of time for the maximally non-Markovian
dephasing dynamics defined by Eq.~(\protect\ref{LindbladDispersivo}).}
\end{figure}
%figura%figura%figura%figura%figurav%figura%figura%figura%figura%figura%figura%figura%figura%figura%figurav%figura%figura%figura%figura%figura
%figura%figura%figura%figura%figurav%figura%figura%figura%figura%figura%figura%figura%figura%figura%figurav%figura%figura%figura%figura%figura

The coherences behavior in Eq. (\ref{solution}) follows straightforwardly as 
$e^{-\Gamma (t)}=[1-2\gamma te^{-\gamma t}].$ From this solution, it is
possible to rewriting the evolution (\ref{LindbladDispersivo}) with a
time-convoluted structure,%
\begin{equation}
\frac{d\rho _{t}}{dt}=\frac{1}{2}\int_{0}^{t}dt^{\prime }k(t-t^{\prime
})(\sigma _{z}\rho _{t^{\prime }}\sigma _{z}-\rho _{t^{\prime }}),
\end{equation}%
with memory kernel%
\begin{equation}
k(t)=2\gamma \lbrack \delta (t)-\gamma \sin (\gamma t)],  \label{DeltaSeno}
\end{equation}%
where $\delta (t)$ is the delta-Dirac function. This kind of kernel, with
both local and nonlocal in time contributions, were also found in Refs. \cite%
{PostMarkovian,eternal}.

\subsection{Random unitary dynamics}

In this case, the qubit dynamics is governed by the time-dependent generator 
$(\sigma _{k}$ is the $k$-Pauli matrix)%
\begin{equation}
\frac{d\rho _{t}}{dt}=\frac{1}{2}\sum_{k=1}^{3}\gamma _{k}(t)(\sigma
_{k}\rho _{t}\sigma _{k}-\rho _{t}),  \label{UnitaryLocalTime}
\end{equation}%
where two decoherence channels were added. The solution is written as a
random unitary map,%
\begin{equation}
\rho _{t}=\sum_{\alpha =0}^{3}p_{\alpha }(t)\sigma _{\alpha }\rho _{0}\sigma
_{\alpha },  \label{UnitaryMap}
\end{equation}%
where $\sigma _{0}=\mathbb{I},$ $0\leq p_{\alpha }(t)\leq 1,$ $\sum_{\alpha
=0}^{3}p_{\alpha }(t)=1.$ The positivity of the weights $\{p_{\alpha }(t)\}$
guarantee the completely positive condition of the solution map.

The set of probabilities$\{p_{\alpha }(t)\}$ and the set of time dependent
rates $\{\gamma _{k}(t)\}$\ depend each of the other. Given the
probabilities, the rates can be expressed as \cite{wuda}%
\begin{equation}
\gamma _{\alpha }(t)=\frac{1}{2}\sum_{\alpha =0}^{3}H_{\alpha \beta }\frac{d%
}{dt}\Big{\{}\ln \Big{[}\sum_{\gamma =0}^{3}H_{\beta \gamma }p_{\gamma }(t)%
\Big{]}\Big{\}},  \label{rates}
\end{equation}%
which implies the relation $\gamma _{0}(t)=-\sum_{k=1}^{3}\gamma _{k}(t).$
The coefficients $\{H_{\alpha \beta }\}$ correspond to a square four
dimensional Hadamard matrix \cite{wuda}. On the other hand, in Appendix B we
find a time-convoluted master equation that is equivalent to Eq.~(\ref%
{UnitaryLocalTime}) (see also Ref. \cite{Admisible}).

Defining $\Gamma _{k}(t)=\int_{0}^{t}\gamma _{k}(\tau )d\tau ,$ the solution
map (\ref{UnitaryMap}) is maximally non-Markovian, for example, if $\gamma
_{2}(t)+\gamma _{3}(t)\ngeq 0$ and \cite{DarioSabrina}%
\begin{equation}
\lim_{t\rightarrow \infty }\Gamma _{1}(t)\geq 0,\ \ \ \ \ \lim_{t\rightarrow
\infty }\Gamma _{2}(t)=\lim_{t\rightarrow \infty }\Gamma _{3}(t)=0.
\label{ConditionsRandomMap}
\end{equation}%
%
%
%
%
%
%
%
%
%
%
%
%
%
%
%
%
%
%
%
%
%
%
%figura1%figura%figura%figura%figurav%figura%figura%figura%figura%figura%figura%figura%figura%figura%figurav%figura%figura%figura%figura%figura
%figura%figura%figura%figura%figurav%figura%figura%figura%figura%figura%figura%figura%figura%figura%figurav%figura%figura%figura%figura%figura
\begin{figure}[tbp]
\includegraphics[bb=13 56 740 600,angle=0,width=7.cm]{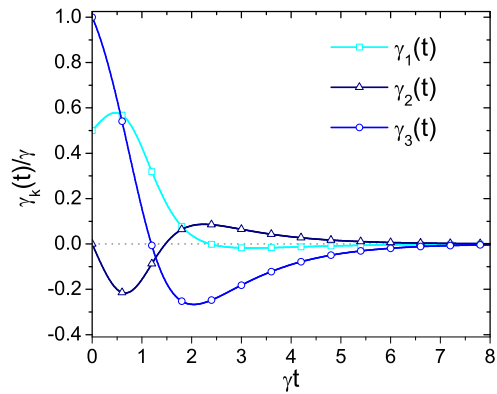}
\caption{Time dependent rates $\{\protect\gamma _{k}(t)\}$ [Eq. (\protect\ref%
{RatesUnitary})] for the maximally non-Markovian random unitary evolution (%
\protect\ref{UnitaryLocalTime}) corresponding to the map (\protect\ref%
{UnitaryMap}) with probabilities (\protect\ref{Probal}).}
\end{figure}
%figura%figura%figura%figura%figurav%figura%figura%figura%figura%figura%figura%figura%figura%figura%figurav%figura%figura%figura%figura%figura
%figura%figura%figura%figura%figurav%figura%figura%figura%figura%figura%figura%figura%figura%figura%figurav%figura%figura%figura%figura%figura

As an explicit example we take the positive and normalized probabilities 
\begin{subequations}
\label{Probal}
\begin{eqnarray}
p_{0}(t) &=&\frac{3}{4}+\frac{1}{4}e^{-\gamma t}\left( 1-2\gamma t\right) ,
\\
p_{1}(t) &=&\frac{1}{4}(1-e^{-\gamma t}),\ \ \ \ \ p_{2}(t)=0, \\
p_{3}(t) &=&\frac{1}{2}\gamma te^{-\gamma t},
\end{eqnarray}%
where $\gamma >0.$ From Eq. (\ref{rates}), the time dependent rates can be
written as 
\end{subequations}
\begin{subequations}
\label{RatesUnitary}
\begin{eqnarray}
\gamma _{1}(t) &=&\frac{1}{2}\gamma \lbrack -g_{a}(t)+g_{b}(t)+g_{c}(t)], \\
\gamma _{2}(t) &=&\frac{1}{2}\gamma \lbrack g_{a}(t)-g_{b}(t)+g_{c}(t)], \\
\gamma _{3}(t) &=&\frac{1}{2}\gamma \lbrack g_{a}(t)+g_{b}(t)-g_{c}(t)],
\end{eqnarray}%
where the auxiliary functions are 
\end{subequations}
\begin{eqnarray}
g_{a}(t) &\equiv &\frac{1-\gamma t}{e^{\gamma t}-\gamma t}, \\
g_{b}(t) &\equiv &\frac{3-2\gamma t}{1+e^{\gamma t}-2\gamma t}, \\
g_{c}(t) &\equiv &\frac{1}{1+e^{\gamma t}}.
\end{eqnarray}%
In Fig. 2 we plot the set $\{\gamma _{k}(t)\}.$ All functions are smooth, do
not diverge, assume positive and negative values, and vanish asymptotically.

From the expressions for the rates [Eq. (\ref{RatesUnitary})] it is possible
to obtain $\lim_{t\rightarrow \infty }\Gamma _{1}(t)=\ln (2),$ while $%
\lim_{t\rightarrow \infty }\Gamma _{2}(t)=\lim_{t\rightarrow \infty }\Gamma
_{3}(t)=0.$ Furthermore, $\gamma _{2}(t)+\gamma _{3}(t)=\gamma g_{a}(t)\ngeq
0.$ Hence, from Eq. (\ref{ConditionsRandomMap}) we conclude that the
evolution is maximally non-Markovian.

\section{Quantum-Classical representation of dynamics}

In the previous section, two maximally non-Markovian evolutions were
explicitly defined. Here, we demonstrate that both examples can be recovered
from simple quantum-classical dynamics, which admit different
representations such as in terms of Markov chains, Lindblad rate equations,
and bipartite evolutions. In all cases, the environment follows its own
unaffected Markovian dynamics. These examples give a basis for constructing
a full class of dynamics with the same memory properties.

\subsection{Dephasing channel}

The dephasing channel is defined by Eq. (\ref{LindbladDispersivo}) with
time-dependent rate (\ref{RateDispersivo}).

\subsubsection{Markov chain representation}

We introduce a classical stochastic system endowed with three different
states (Fig. 3). Their probabilities $\{P_{i}(t)\},$ $i=0,1,2,$ obey by the
classical master equation 
\begin{subequations}
\label{ClasicalMaster}
\begin{eqnarray}
\frac{dP_{0}(t)}{dt} &=&-\gamma P_{0}(t), \\
\frac{dP_{1}(t)}{dt} &=&-\gamma P_{1}(t)+\gamma P_{0}(t), \\
\frac{dP_{2}(t)}{dt} &=&+\gamma P_{1}(t),
\end{eqnarray}%
with initial condition $P_{0}(0)=1.$ Thus, at random times the classical
system undergoes the successive transitions $0\rightarrow 1\rightarrow 2.$
In addition, in each transition the quantum system suffers the disruptive
transformation 
\end{subequations}
\begin{equation}
\rho \rightarrow \sigma _{z}\rho \sigma _{z}.  \label{Collisional}
\end{equation}%
Therefore, the density matrix of the quantum system is 
\begin{equation}
\rho _{t}=P_{0}(t)\rho _{0}+P_{1}(t)\sigma _{z}\rho _{0}\sigma
_{z}+P_{2}(t)\sigma _{z}^{2}\rho _{0}\sigma _{z}^{2}.  \label{ChainRho}
\end{equation}%
This solution can be read as a quantum collisional model \cite{collisional}
with a finite number of events.

Eq. (\ref{ChainRho}) can trivially be written with the structure given by
Eq. (\ref{UnitaryMap}), with $p_{0}(t)=P_{0}(t)+P_{2}(t),$ $%
p_{1}(t)=p_{2}(t)=0,$ and $p_{3}(t)=P_{1}(t).$ By solving Eq.~(\ref%
{ClasicalMaster}) with $P_{0}(0)=1,$ we get%
\begin{equation}
p_{0}(t)=1-\gamma t\exp (-\gamma t),\ \ \ \ \ \ p_{3}(t)=\gamma t\exp
(-\gamma t).  \label{pminuscula}
\end{equation}%
These expressions in turn, from Eq. (\ref{rates}), lead to the time
dependent rate (\ref{RateDispersivo}), which recovers in consequence the
maximally non-Markovian quantum dephasing evolution introduced previously.
This result can alternatively be derived from Eq. (\ref{ChainRho}) by
obtaining the coherences time behavior. 
%figura1%figura%figura%figura%figurav%figura%figura%figura%figura%figura%figura%figura%figura%figura%figurav%figura%figura%figura%figura%figura
%figura%figura%figura%figura%figurav%figura%figura%figura%figura%figura%figura%figura%figura%figura%figurav%figura%figura%figura%figura%figura
\begin{figure}[tbp]
\includegraphics[bb=80 215 540 325,angle=0,width=7.5cm]{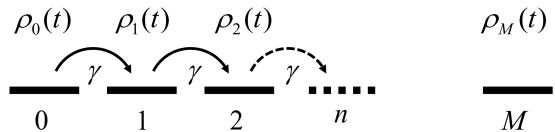}
\caption{Graphical representation of the quantum-classical hybrid dynamics.
For the dephasing channel [Eqs.~(\protect\ref{LindbladDispersivo}) and (%
\protect\ref{RateDispersivo})], the environment consists in three
(unidirectionally) coupled states, which obey the classical master equation (%
\protect\ref{ClasicalMaster}). In each transition, the transformation (%
\protect\ref{Collisional}) is applied. The quantum state follows from the
Lindblad rate evolution (\protect\ref{LindbladRate}). Extra coupled states
(dashed line) may lead to decoherence and recoherence, Eq.~(\protect\ref%
{Recoherence}). A single uncoupled state $M$ completes the scheme for the
random unitary evolution [Eqs. (\protect\ref{UnitaryLocalTime}) and (\protect
\ref{RatesUnitary})], whose solution is the mixed state (\protect\ref%
{RhoMezcla}).}
\end{figure}
%figura%figura%figura%figura%figurav%figura%figura%figura%figura%figura%figura%figura%figura%figura%figurav%figura%figura%figura%figura%figura
%figura%figura%figura%figura%figurav%figura%figura%figura%figura%figura%figura%figura%figura%figura%figurav%figura%figura%figura%figura%figura

In the previous stochastic representation, the property $\lim_{t\rightarrow
\infty }\rho _{t}=\rho _{0}$ can be understood straightforwardly from the
occurrence of two successive flips in the sign of the coherences, which are
induced by the transformation $\rho \rightarrow \sigma _{z}\rho \sigma _{z}.$
Furthermore, the environment, represented here by the classical system, is
completely unaware of the (unidirectional) dependence of the system dynamics
on its states.

\subsubsection{Lindblad rate equation}

As is well known, the class of stochastic dynamics described previously can
be formulated in terms of Lindblad rate equations \cite{LindbladRate}. These
generalized Lindblad equations give the more general evolution for a
quantum-classical (hybrid) arrange where the behavior of the classical part
is inherently irreversible in time.

The system density matrix is written as%
\begin{equation}
\rho _{t}=\sum_{i=0}^{2}\rho _{i}(t),  \label{solutionLRate}
\end{equation}%
where the auxiliary states $\{\rho _{i}(t)\}$\ evolve as 
\begin{subequations}
\label{LindbladRate}
\begin{eqnarray}
\frac{d\rho _{0}(t)}{dt} &=&-\gamma \rho _{0}(t), \\
\frac{d\rho _{1}(t)}{dt} &=&-\gamma \rho _{1}(t)+\gamma \sigma _{z}\rho
_{0}(t)\sigma _{z}, \\
\frac{d\rho _{2}(t)}{dt} &=&+\gamma \sigma _{z}\rho _{1}(t)\sigma _{z},
\end{eqnarray}%
with initial conditions $\rho _{0}(0)=\rho _{0}$ and $\rho _{1}(0)=\rho
_{2}(0)=0.$ This Lindblad rate equation admits the stochastic representation
defined previously. In fact, here the evolutions for the classical
populations [Eq. (\ref{ClasicalMaster})] are recovered from $P_{i}(t)=%
\mathrm{Tr}[\rho _{i}(t)],$ $i=0,1,2.$ On the other hand, the
transformations $\rho \rightarrow \sigma _{z}\rho \sigma _{z}$ are taken
into account in Eq.~(\ref{LindbladRate}) through the coupling between the
auxiliary states. The explicit solution for the system density matrix (\ref%
{solutionLRate}) recovers the result Eq. (\ref{solution}).

Lindblad rate equations also arise from a generalized Born-Markov
approximation. Hence, the previous dynamics [Eq. (\ref{LindbladRate})] can
alternatively be read as the result of the interaction with a complex
structured reservoir~\cite{LindbladRate}.

\subsubsection{Bipartite Lindblad representation and measurement trajectories%
}

Lindblad rate equations can be embedded in a bipartite quantum dynamics \cite%
{LindbladRate}. In the present case, the \textquotedblleft ancilla
system\textquotedblright\ has three states, $|0\rangle ,$ $|1\rangle ,$ and $%
|2\rangle .$ Denoting with $\rho _{t}^{sa}$ the system and ancilla density
matrix, its evolution can be written as 
\end{subequations}
\begin{eqnarray}
\frac{d\rho _{t}^{sa}}{dt} &=&+\frac{\gamma }{2}([V_{1},\rho
_{t}^{sa}V_{1}^{\dag }]+[V_{1}\rho _{t}^{sa},V_{1}^{\dag }])  \notag \\
&&+\frac{\gamma }{2}([V_{2},\rho _{t}^{sa}V_{2}^{\dag }]+[V_{2}\rho
_{t}^{sa},V_{2}^{\dag }]),  \label{BipartiteLindblad}
\end{eqnarray}%
where the initial state is $\rho _{0}^{sa}=\rho _{0}\otimes |0\rangle
\langle 0|.$ The dissipative channels are defined by the operators%
\begin{equation}
V_{1}=\sigma _{z}\otimes |1\rangle \langle 0|,\ \ \ \ \ \ V_{2}=\sigma
_{z}\otimes |2\rangle \langle 1|.
\end{equation}%
The dynamics given by the Lindblad rate equation (\ref{LindbladRate}) is
recovered from $\rho _{i}(t)=\langle i|\rho _{t}^{sa}|i\rangle ,$ $i=0,1,2.$

The bipartite representation [Eq. (\ref{BipartiteLindblad})], through the
standard quantum jump approach \cite{breuerBook}, allows to unravelling the
evolution in measurement trajectories. Assuming that the measurement device
detects the transitions $|0\rangle \rightarrow |1\rangle $ and $|1\rangle
\rightarrow |2\rangle ,$ the\ bipartite measurement operators are $V_{1}$\
and $V_{2}.$\ Similarly to Ref. \cite{embedding}, the \textquotedblleft
collisional trajectories,\textquotedblright\ defined by Eq. (\ref%
{Collisional}), can be recovered from the bipartite measurement
realizations. The main difference is that here only a finite number of
transitions occur.

\subsubsection{Decoherence and recoherence}

The underlying quantum-classical dynamics can be generalized without
affecting the non-Markovianity degree of the solution map. For example, the
introduction of extra classical states also leads to the property $\lim
{}_{t\rightarrow \infty }\rho _{t}=\rho _{0}$ if an even number of
transformations~(\ref{Collisional}) happen (Fig. 3). By increasing the
number of collisions\ the phenomenon of \textit{recoherence} occurs, that
is, after an exponential decay the system coherence (almost) vanishes and
emerges at a later time, recovering its initial value in the (stationary)
long time regime.

Denoting the qubit coherence as $\rho _{12}(t)=c_{n}(t)\rho _{12}(0),$ where
the function $c_{n}(t)$ gives its characteristic time behavior and the
subindex $n$ denotes the (even) number of collisions, in Appendix C we obtain%
\begin{eqnarray}
c_{n}(t) &=&e^{-2\gamma t}+\int_{0}^{t}dt_{2}\int_{0}^{t_{2}}dt_{1}2\gamma
e^{-2\gamma (t_{2}-t_{1})}  \notag \\
&&\times \left[ \frac{\gamma e^{-\gamma t_{1}}}{(n-1)!}(\gamma t_{1})^{n-1}%
\right] .  \label{Recoherence}
\end{eqnarray}%
This expression follows by conditioning the quantum dynamics to a classical
system with $n+1$ states, where each transition implies the transformation (%
\ref{Collisional}). The (unidirectional) coupling rate between successive
states is $\gamma .$ For the model with three states (Fig.~1), it follows $%
c_{2}(t)=1-2\gamma te^{-\gamma t}=e^{-\Gamma (t)},$ where consistently $%
\Gamma (t)$ is given by Eq. (\ref{GamaDispersivo}). In all cases the time
dependent rate [Eq. (\ref{LindbladDispersivo})] follows as $\gamma
(t)=(d/dt)\ln [1/c_{n}(t)],$ with $\Gamma (t)=\ln [1/c_{n}(t)],$ which
fulfills Eq. (\ref{condition}). 
%figura1%figura%figura%figura%figurav%figura%figura%figura%figura%figura%figura%figura%figura%figura%figurav%figura%figura%figura%figura%figura
%figura%figura%figura%figura%figurav%figura%figura%figura%figura%figura%figura%figura%figura%figura%figurav%figura%figura%figura%figura%figura
\begin{figure}[tbp]
\includegraphics[bb=13 56 740 600,angle=0,width=7.cm]{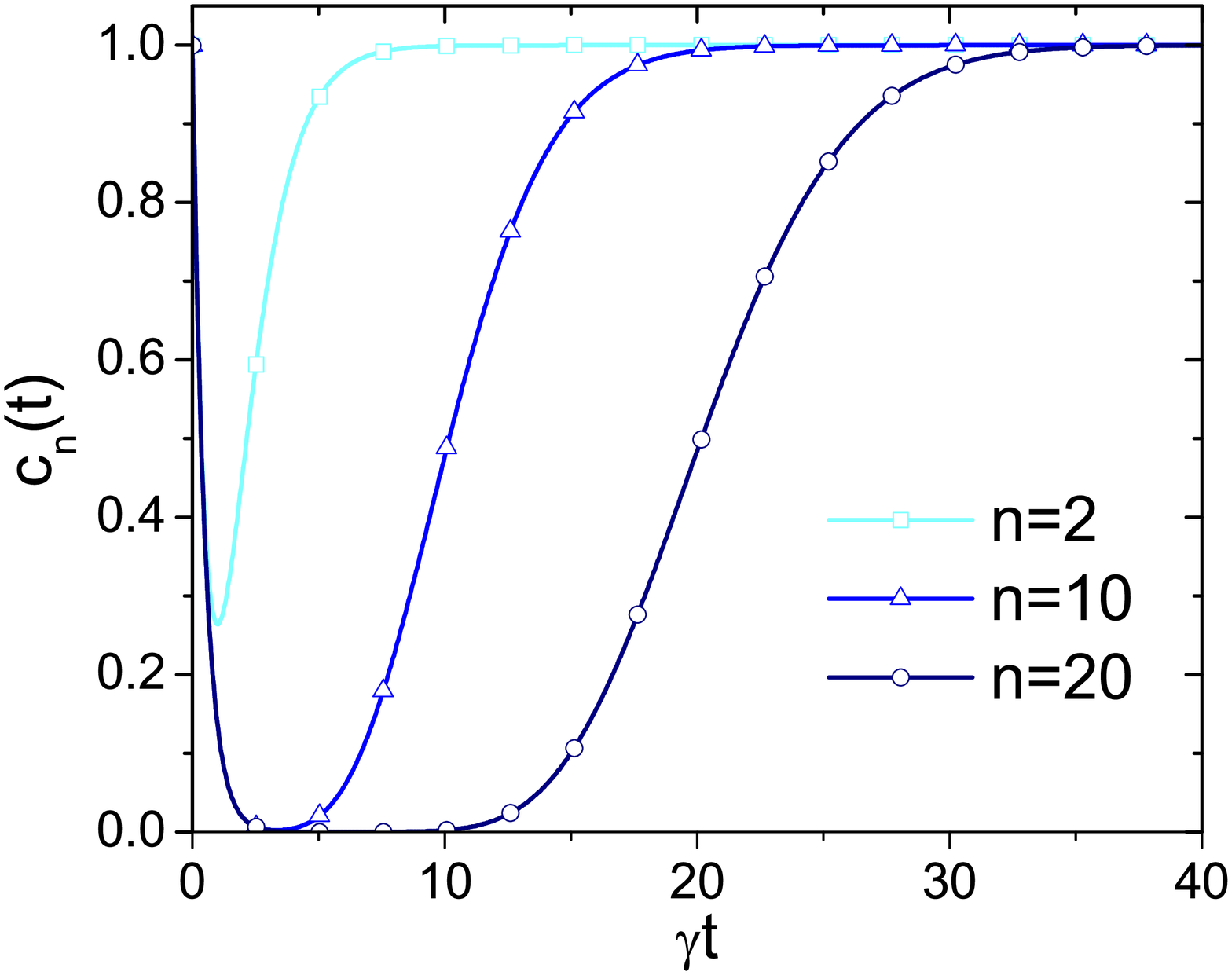}
\caption{Time dependence $c_{n}(t)$ of the qubit coherence for the extended
dephasing model, Eq.~(\protect\ref{Recoherence}), for different (even)
number of collisional events, $n=2,$ $10,$ and $20.$}
\end{figure}
%figura%figura%figura%figura%figurav%figura%figura%figura%figura%figura%figura%figura%figura%figura%figurav%figura%figura%figura%figura%figura
%figura%figura%figura%figura%figurav%figura%figura%figura%figura%figura%figura%figura%figura%figura%figurav%figura%figura%figura%figura%figura

In Fig. 4 it is plotted $c_{n}(t)$ as a function of time for $n=2,$ $10,$
and $20.$ For $n\gg 1,$ the coherence vanishes following the decay behavior $%
c_{n}(t)\simeq \exp (-2\gamma t),$ while recoherence always occurs in the
long time regime, $\lim_{t\rightarrow \infty }c_{n}(t)=1.$ On the other
hand, a Markovian limit is recovered as $\lim_{n\rightarrow \infty
}c_{n}(t)=\exp (-2\gamma t).$ In this case, a standard collisional dynamics
is recovered \cite{collisional}, which in a given realization contains an
infinite number of disruptive transformations.

\subsection{Random unitary dynamics}

The Maximally non-Markovian random unitary dynamics defined by Eq. (\ref%
{UnitaryMap}) and the probabilities (\ref{Probal}) can also be derived
taking into account an environment that is unaffected by the system
dynamics. With this goal, in addition to the three auxiliary states defined
by Eq.~(\ref{LindbladRate}), an extra state $\rho _{M}(t)$ is introduced
(Fig. 3). It is completely uncoupled of the other states and follows the
Markovian evolution%
\begin{equation}
\frac{d\rho _{M}(t)}{dt}=\frac{1}{2}\gamma \lbrack \sigma _{x}\rho
_{M}(t)\sigma _{x}-\rho _{M}(t)].  \label{LindbladM}
\end{equation}%
Here, $\sigma _{x}$ is the $x-$Pauli matrix. The solution is given by%
\begin{equation}
\rho _{M}(t)=\frac{1}{2}[(1+e^{-\gamma t})\rho _{0}+(1-e^{-\gamma t})\sigma
_{x}\rho _{0}\sigma _{x}].  \label{MarkovSolution}
\end{equation}%
The system density matrix is written as%
\begin{equation}
\rho _{t}=(1-r)\sum_{i=0}^{2}\rho _{i}(t)+r\rho _{M}(t),  \label{RhoMezcla}
\end{equation}%
where $0\leq r\leq 1.$ This expression corresponds to a statistical mixture
(or convex combination \cite{Convex}) of a Markovian and a maximally
non-Markovian dephasing evolution, Eqs. (\ref{LindbladM}) and (\ref%
{LindbladDispersivo}) respectively. Taking $r=1/2,$ from the solutions (\ref%
{ChainRho}) and (\ref{MarkovSolution}) it follows the probabilities (\ref%
{Probal}), recovering in consequence the maximally non-Markovian random
unitary evolution presented in the previous section. Notice that also in
this case the environment is unaware of the dependence of the system
dynamics on its states, implying in consequence, the absence of any physical
backflow of information. On the other hand, the state (\ref{RhoMezcla}) can
also be characterized in terms of Lindblad rate equations and bipartite
evolutions.

It is interesting to note that Eq. (\ref{RhoMezcla}) leads to a Markovian
solution map for $r=1.$ By an explicit calculation, it is possible to
demonstrate that this is the unique value at which the map departs from a
maximally non-Markovian dynamics. In fact, for $0\leq r<1$ we get $\gamma
_{2}(t)+\gamma _{3}(t)=\gamma g_{a}(r,t)\ngeq 0,$ $\lim_{t\rightarrow \infty
}\Gamma _{1}(t)=\ln [1/(1-r)],$\ and $\lim_{t\rightarrow \infty }\Gamma
_{2}(t)=\lim_{t\rightarrow \infty }\Gamma _{3}(t)=0,$ where $%
g_{a}(r,t)=[2(1-r)(1-\gamma t)]/[e^{\gamma t}-2(1-r)\gamma t].$ Hence,
condition Eq.~(\ref{ConditionsRandomMap}) is fulfilled. In contrast, a
statistical mixture of a maximally entangled state and the identity matrix
state (Werner state) departs from an entangled state at $r<1$ \cite%
{horodecki}. This difference is expectable because the non-Markovianity
degree relies on a formal analogy with entanglement \cite{DarioSabrina}.

\section{Non-Markovian quantum dynamics without environment backflow of
information}

The previous examples, as well as the results of Refs.~\cite%
{PostMarkovian,embedding,eternal}, demonstrate that strong non-Markovian
effects may arise from simple quantum-classical hybrid dynamics. The
environment consists in a set of classical degrees of freedom that follows
their own dynamics. Here, we write the more general underlying equations
that lead to these properties and analyze how memory effects may emerge.

\subsection{Discrete dissipative classical degrees of freedom}

The system density matrix is written in terms of a set of auxiliary states, $%
\rho _{t}=\sum_{R}\rho _{R}(t),$ whose evolution is given by a Lindblad rate
equation,%
\begin{eqnarray}
\frac{d\rho _{R}(t)}{dt} &=&\frac{-i}{\hbar }[H_{R},\rho _{R}(t)]+\mathbb{L}%
_{R}[\rho _{R}(t)]  \label{LindbladRateConE} \\
&&-\sum_{R^{\prime }}\gamma _{R^{\prime }R}\rho _{R}(t)+\sum_{R^{\prime
}}\gamma _{RR^{\prime }}\mathbb{E}_{RR^{\prime }}[\rho _{R^{\prime }}(t)], 
\notag
\end{eqnarray}%
with initial conditions $\rho _{R}(0)=P_{R}(0)\rho _{0}.$ In the first line, 
$H_{R}$ and $\mathbb{L}_{R}$ are the (conditional) unitary and dissipative
Lindblad dynamics of the system given that the environment is in the state $%
R.$ The second line introduces a coupling between the auxiliary states with
rates $\{\gamma _{R^{\prime }R}\}.$ In each transition $R^{\prime
}\rightarrow R,$ the state $\rho _{R^{\prime }}$ suffers the disruptive
change $\mathbb{E}_{RR^{\prime }}[\rho _{R^{\prime }}]\rightarrow \rho _{R},$
where $\{\mathbb{E}_{RR^{\prime }}[\bullet ]\}$ are trace preserving
completely positive superoperators,%
\begin{equation}
\mathrm{Tr}(\mathbb{E}_{RR^{\prime }}[\rho ])=\mathrm{Tr}(\rho ).
\end{equation}

The probability that the environment is in the state $R$ at time $t$ is
given by $P_{R}(t)=\mathrm{Tr}[\rho _{R}(t)]$ \cite{LindbladRate}. From Eq.~(%
\ref{LindbladRateConE}), it follows the classical master equation%
\begin{equation}
\frac{dP_{R}(t)}{dt}=-\sum_{R^{\prime }}\gamma _{R^{\prime
}R}P_{R}(t)+\sum_{R^{\prime }}\gamma _{RR^{\prime }}P_{R}^{\prime }(t).
\label{MasterClassicalGen}
\end{equation}%
This result implies that, in fact, the environment is completely unperturbed
by the system state or dynamics. Hence, the environment is only a
\textquotedblleft spectator\textquotedblright\ or \textquotedblleft casual
bystander\textquotedblright\ during the whole evolution \cite{eternal}. Eq.~(%
\ref{LindbladRateConE}) is the more general quantum-classical evolution that
leads to these properties \cite{demo}.

The underlying equations (\ref{LindbladRateConE}) may induce strong
non-Markovian quantum effects, which in turn may depend on the initials
conditions $\{P_{R}(0)\}.$ For example, the particular case (\ref%
{LindbladRate}) is recovered with $H_{R}\rightarrow 0,$ $\mathbb{L}%
_{R}\rightarrow 0,$ and $\mathbb{E}_{RR^{\prime }}[\bullet ]\rightarrow
\sigma _{z}\bullet \sigma _{z}$ \cite{initials}.

The possibility of observing memory or non-Markovian effects without a
physical environment-to-system backflow of information follows from the 
\textit{unidirectional dependence of the system evolution on the environment
states} [Eq.~(\ref{LindbladRateConE})]. This dependence is present even when 
$\mathbb{E}_{RR^{\prime }}=\mathbb{I}$ or when $\{\gamma _{R^{\prime
}R}\}=0. $ In this last extreme case, $\{P_{R}(t)\}=\{P_{R}(0)\}.$ By
writing the system density matrix as $\rho _{t}=\Lambda _{t}\rho
_{0}=V_{t,s}\Lambda _{t}\rho _{0},$ we realize that the intermediate
propagator $V_{t,s}$ (which is used for defining the non-Markovianity degree 
\cite{DarioSabrina}) may develops strong departures from a completely
positive \textquotedblleft Markovian\textquotedblright\ propagator. In fact, 
$V_{t,s}$ must to takes into account the previous dependence of the system
dynamics on the environment states. This fact explain why, \textit{even when
the environment is frozen} $\{P_{R}(t)\}=\{P_{R}(0)\}$ \cite{eternal}, the
system dynamics may develops memory effects without happening a physical
environment-to-system backflow of information.

\subsection{Non-white Gaussian stochastic Hamiltonian}

Non-Markovian effects induced by non-white Gaussian stochastic Hamiltonians 
\cite{GaussianNoise}\ also falls in the previous category. In order to show
this fact, we write the density matrix as%
\begin{equation}
\rho _{t}=\int_{-\infty }^{+\infty }d\eta \rho _{\eta }(t),  \label{RhoNoise}
\end{equation}%
where the auxiliary state $\rho _{\eta }(t)$ depends on the real parameter $%
\eta ,$ and obeys the equation%
\begin{eqnarray}
\frac{\partial \rho _{\eta }(t)}{\partial t} &=&\frac{-i}{\hbar }[H+\eta
\Delta H,\rho _{\eta }(t)]  \label{LindbladRateNoise} \\
&&-\gamma \frac{\partial \lbrack \eta \rho _{\eta }(t)]}{\partial \eta }+%
\frac{D}{2}\frac{\partial ^{2}\rho _{\eta }(t)}{\partial \eta ^{2}}.  \notag
\end{eqnarray}%
Both $H$ and $\Delta H$ are Hamiltonian operators. The environment dynamics,
defined by the variable $\eta ,$ follows from $P_{\eta }(t)\equiv \mathrm{Tr}%
[\rho _{\eta }(t)],$ which gives%
\begin{equation}
\frac{\partial P_{\eta }(t)}{\partial t}=-\gamma \frac{\partial \lbrack \eta
P_{\eta }(t)]}{\partial \eta }+\frac{D}{2}\frac{\partial ^{2}P_{\eta }(t)}{%
\partial \eta ^{2}}.  \label{FP}
\end{equation}%
This Fokker-Planck equation \cite{vanKampen} is independent of the system
state. Furthermore, the quantum state parametrically depends on the bath
state $\eta .$ Therefore, non-Markovian effects obtained from Eq. (\ref%
{LindbladRateNoise}) occur without a physical environment-to-system backflow
of information. On the other hand, this equation can be mapped to a
stochastic Hamiltonian evolution. In fact, Eq. (\ref{FP}) is equivalent to
the Langevin equation~\cite{vanKampen}%
\begin{equation}
\frac{d}{dt}\eta (t)=-\gamma \eta (t)+\xi (t),
\end{equation}%
where the white Gaussian noise $\xi (t)$ has correlation $\langle \xi (t)\xi
(t^{\prime })\rangle =D\delta (t-t^{\prime }).$ In contrast, the non-white
Gaussian process $\eta (t)$ has an exponential correlation, which in the
stationary regime reads $\langle \eta (t)\eta (t^{\prime })\rangle _{\mathrm{%
st}}=D/(2\gamma )\exp (-\gamma |t-t^{\prime }|).$ In addition, the system
density matrix $\rho _{t}$ [Eq. (\ref{RhoNoise})] can be read as an average
over realizations of $\eta (t),$ $\rho _{t}=\langle \rho _{st}(t)\rangle ,$
where $\rho _{st}(t)$ follows the stochastic evolution%
\begin{equation}
\frac{d\rho _{st}(t)}{dt}=\frac{-i}{\hbar }[H+\eta (t)\Delta H,\rho
_{st}(t)].
\end{equation}%
Therefore, stochastic non-white Hamiltonians also lead to non-Markovian
effects \cite{GaussianNoise}\ without happening a physical
environment-to-system backflow of information.

\section{Summary and conclusions}

Similarly to Schmidt number in entanglement theory, the degree of
non-Markovianity allows to ranking memory effects developed by quantum
evolutions. Maximally non-Markovian evolutions are the analog of maximally
entangled states. Here, we showed that exist a class of such kind of extreme
non-Markovian dynamics that can be modeled from simple quantum-classical
hybrid dynamics. The environment, defined by the classical degrees of
freedom, follows its own Markovian evolution, which in turn is unaffected by
the system state or dynamics. Therefore, memory effects arise without the
occurrence of any physical environment-to-system backflow of information.
Departure with respect to a quantum Markovian regime follows from the
unidirectional dependence of the system dynamics on the environment states.

Both dephasing and random unitary maximally non-Markovian maps were modeled.
In the former case, the dynamics can be read in terms of a finite (even)
number of collisional events that change the sign of the qubit coherence,
each event occurring at random times. The phenomena of decoherence and
recoherence are captured in this frame. On the other hand, in the last case
(random unitary maps), the system dynamics follows from a statistical
mixture (convex combination) of the former non-Markovian evolution and a
Markovian dephasing channel in a different direction. In both cases, the
classical degrees of freedom are unaffected by the system state. The
underlying dynamics admit different representations such as in terms of
quantum Markov chains, Lindblad rate equations, and measurement trajectories.

General Lindblad rate equations that lead to memory effects without a
physical backflow of information were also characterized. The proposed
models can be generalized in different ways (dimensionality, collision
transformations, quantum ancillas, etc.), providing a basis for the study of
a full family of maximally non-Markovian dynamics. This approach may be
useful for modeling decoherence in quantum information channels where the
environment is coupled during a set of finite intervals of time (collisional
models with a finite number of events).

In accordance with previous results \cite{PostMarkovian,embedding,eternal},
the present analyses confirm that, in general, non-Markovian indicators or
measures based solely on the system dynamics are \textit{unable to
distinguish} between memory (non-Markovian) effects developed with or
without a physical environment-to-system backflow of information.

\section*{Acknowledgments}

This work was supported by Consejo Nacional de Investigaciones Cient\'{\i}%
ficas y T\'{e}cnicas (CONICET), Argentina.

\appendix

\section{Non-Markovianity degree}

Here, we briefly review the formalism of Ref.~\cite{DarioSabrina}. It is
based on the divisibility of the density matrix propagator, $\rho
_{t}=\Lambda _{t}\rho _{0}.$ Splitting the time evolution as $\Lambda
_{t}=V_{t,s}\Lambda _{s},$\ the solution map $\Lambda _{t}$\ is called $k-$%
\textit{divisible} if $V_{t,s}$ is $k-$positive for all $t\geq s\geq 0,$
that is, $\mathbb{I}_{k}\otimes V_{t,s}$ is positive, where $\mathbb{I}_{k}$
is the identity operator in an extra Hilbert space of dimension $k.$

The \textit{degree of non-Markovianity} is a number associated to a given
evolution. The solution map $\Lambda _{t}$\ has a non-Markovianity degree $%
\mathrm{NMD}[\Lambda _{t}]=k$ if and only if $\Lambda _{t}$\ is $(n-k)$ but
not $(n+1-k)$ divisible. $\Lambda _{t}$ is Markovian if and only if $\mathrm{%
NMD}[\Lambda _{t}]=0$ and essentially non-Markovian if and only if $\mathrm{%
NMD}[\Lambda _{t}]=n,$ where $n$ is the system Hilbert space dimension.

If $\Lambda _{t}$\ is $k-$divisible, then $(d/dt)||\mathbb{I}_{k}\otimes
\Lambda _{t}||_{1}\leq 0,$ for all operators $X$ in the bipartite Hilbert
space of the system and ancilla of dimension $k.$ This result allows to
defining a series of \textit{non-Markovianity measures}. Departure from $k$
divisibility is measured with%
\begin{equation}
\mathcal{M}_{k}[\Lambda _{t}]=\sup_{X}\frac{N_{k}^{+}[X]}{|N_{k}^{-}[X]|},
\end{equation}%
where%
\begin{equation}
N_{k}^{+}[X]=\int_{\lambda _{k}(X;t)>0}\lambda _{k}(X;t)dt,
\end{equation}%
and, similarly for $N_{k}^{+}[X]$ (where the integral is over intervals such
that $\lambda _{k}(X;t)<0),$ and%
\begin{equation}
\lambda _{k}(X;t)=\frac{d}{dt}||\mathbb{I}_{k}\otimes \Lambda _{t}||_{1}.
\end{equation}%
The supremum is taken over all Hermitian $X.$ Using this expression it
follows $\int_{0}^{\infty }\lambda _{k}(X;t)dt\leq 0,$ which proves that $%
|N_{k}^{-}[X]|\geq N_{k}^{+}[X],$ and in consequence $\mathcal{M}%
_{k}[\Lambda _{t}]\in \lbrack 0,1].$ If $l>k,$ then $\mathcal{M}_{l}[\Lambda
_{t}]\geq \mathcal{M}_{k}[\Lambda _{t}]$ and, hence%
\begin{equation}
0\leq \mathcal{M}_{1}[\Lambda _{t}]\leq \cdots \leq \mathcal{M}_{n}[\Lambda
_{t}]\leq 1.
\end{equation}%
This expression is the analog of a similar relation between the coefficients
of a Schmidt decomposition of an entangled pure quantum state. \textit{%
Maximally non-Markovian dynamics} corresponds to the case $\mathcal{M}%
_{1}[\Lambda _{t}]=1,$ which immediately leads to%
\begin{equation}
\mathcal{M}_{1}[\Lambda _{t}]=\cdots =\mathcal{M}_{n}[\Lambda _{t}]=1,
\end{equation}%
in perfect analogy with maximally quantum entangled states \cite{horodecki}.

The conditions under which the maps (\ref{LindbladDispersivo}) and (\ref%
{UnitaryLocalTime}) are maximally non-Markovian [Eqs. (\ref{condition}) and (%
\ref{ConditionsRandomMap}) respectively]\ were derived in Ref.~\cite%
{DarioSabrina} from the condition $\mathcal{M}_{1}[\Lambda _{t}]=1,$ and
using, for example, the operator $X=\sigma _{x}.$

\section{Time-convoluted evolution for random unitary maps}

Given the evolution defined by the general random unitary map Eq. (\ref%
{UnitaryMap}), the goal here is to obtain a time-convoluted evolution%
\begin{equation}
\frac{d}{dt}\rho _{t}=\int_{0}^{t}d\tau \mathbb{L}_{t-\tau }\rho _{\tau },
\end{equation}%
where $\mathbb{L}_{t}$ is a memory superoperator. The derivation is similar
to the time-convolutionless case \cite{wuda}. Expressing $\rho _{t}$ in
terms of its propagator,\ $\rho _{t}=\Lambda _{t}\rho ,$ and working the
previous time evolution in the Laplace domain $[f(z)=\int_{0}^{\infty
}dte^{-zt}f(t)],$ it follows the relation%
\begin{equation}
\mathbb{L}_{z}=(z\Lambda _{z}-1)\frac{1}{\Lambda _{z}}.  \label{Lz}
\end{equation}%
Writing Eq. (\ref{UnitaryMap}) in the Laplace domain,%
\begin{equation}
\Lambda _{z}\rho =\sum_{\alpha =0}^{3}p_{\alpha }(z)\sigma _{\alpha }\rho
\sigma _{\alpha },
\end{equation}%
similarly to Ref. \cite{wuda}, let us observe that%
\begin{equation}
\Lambda _{z}(\sigma _{\alpha })=\lambda _{\alpha }(z)\sigma _{\alpha },
\label{propaMemo}
\end{equation}%
where the eigenvalues are given by%
\begin{equation}
\lambda _{\alpha }(z)=\sum_{\beta =0}^{3}H_{\alpha \beta }p_{\beta }(z),
\end{equation}%
with $H_{\alpha \beta }$ being a square Hadamard matrix $H=\{\{1,1,1,1\},%
\{1,1,-1,-1\},\{1,-1,1,-1\},\{1,-1,-1,1\}\}$ \cite{wuda}. Notice that $%
\lambda _{0}(z)=1/z.$

From Eqs. (\ref{Lz}) and (\ref{propaMemo}) it follows that%
\begin{equation}
\mathbb{L}_{z}(\sigma _{\alpha })=\mu _{\alpha }(z)\sigma _{\alpha },
\label{LatSigma}
\end{equation}%
where%
\begin{equation}
\mu _{\alpha }(z)=\frac{[z\lambda _{\alpha }(z)-1]}{\lambda _{\alpha }(z)}.
\label{mus}
\end{equation}%
Assuming that $\mathbb{L}_{z}$ can be written as%
\begin{equation}
\mathbb{L}_{z}(\rho )=\frac{1}{2}\sum_{\alpha =0}^{3}k_{\alpha }(z)\sigma
_{\alpha }\rho \sigma _{\alpha },
\end{equation}%
where $\{k_{\alpha }(z)\}$ are memory functions, using the property (\ref%
{propaMemo}) with $p_{\alpha }(z)\rightarrow k_{\alpha }(z),$ it follows
that $\mathbb{L}_{z}(\sigma _{\alpha })=(1/2)\sum_{\alpha =0}^{3}H_{\alpha
\beta }k_{\beta }(z)\sigma _{\alpha }.$ From this relation and Eq.~(\ref%
{LatSigma}), and by using that $H^{-1}=(1/4)H,$ it follows a close
expression for the memory functions%
\begin{equation}
k_{\alpha }(z)=\frac{1}{2}\sum_{\beta =0}^{3}H_{\alpha \beta }\mu _{\beta
}(z).  \label{kernel}
\end{equation}%
Finally, by using that the sum $\sum_{\beta =0}^{3}k_{\alpha
}(z)=(1/2)\sum_{\beta =0}^{3}H_{\alpha \beta }\mu _{\beta }(z)=\mu _{0}(z)=0,
$ if follows the final standard form%
\begin{equation}
\mathbb{L}_{z}(\rho )=\frac{1}{2}\sum_{j=1}^{3}k_{j}(z)(\sigma _{j}\rho
\sigma _{j}-\rho ).
\end{equation}%
The memory functions $\{k_{\alpha }(z)\}$ only depend on the set of
probabilities $\{p_{\alpha }(z)\}.$ On the other hand, general kernels $%
\{k_{\alpha }(z)\}$ that guarantee a completely positive solution map were
characterized in Ref. \cite{Admisible}.

For the dephasing map defined by Eqs. (\ref{LindbladDispersivo}) and (\ref%
{RateDispersivo}), the probabilities are $p_{0}(t)=1-\gamma t\exp (-\gamma
t),$ $p_{1}(t)=p_{2}(t)=0,$ and $p_{3}(t)=\gamma t\exp (-\gamma t),$ Eq.~(%
\ref{pminuscula}). From Eq. (\ref{kernel}) it follows $k_{1}(z)=k_{2}(z)=0,$
and $k_{3}(z)=2z^{2}\gamma /(z^{2}+\gamma ^{2}),$ which consistently is the
Laplace transform of Eq.~(\ref{DeltaSeno}). For the probabilities Eq. (\ref%
{Probal}), the kernels have the same structure, where their specific time
dependence can be written in terms of exponential and trigonometric
functions.

\section{Decoherence and recoherence calculus}

Here, we derive the coherence behavior for a collisional dynamics where $n$%
-applications of the disruptive transformation $\rho \rightarrow \sigma
_{z}\rho \sigma _{z}$ are applied at random times.\ The probability density
(waiting function) for the consecutive intervals is $w_{t}=\gamma \exp
(-\gamma t),$ which has associated the survival probability $%
s_{t}=1-\int_{0}^{t}w_{t^{\prime }}dt^{\prime }=\exp (-\gamma t).$ The
following derivation is valid for any $w_{t}.$

The coherence is written as $\rho _{12}(t)=c_{n}(t)\rho _{12}(0).$ Given
that an \textit{even} finite number of collisions $(n)$ happen, that each
collision changes the sign of the coherence, in the Laplace domain $%
[f(z)=\int_{0}^{\infty }dte^{-zt}f(t)]$ it follows%
\begin{eqnarray}
c_{n}(z) &=&s_{z}(1+w_{z}^{2}+w_{z}^{4}+\cdots w_{z}^{n-2})+\frac{w_{z}^{n}}{%
z}  \notag \\
&&-s_{z}(w_{z}+w_{z}^{3}+\cdots w_{z}^{n-1}),
\end{eqnarray}%
where even and odd numbers of intermediate collisions were taken into
account. Rewriting the previous expression as%
\begin{equation}
c_{n}(z)=s_{z}(1-w_{z})\sum_{i=0}^{\frac{n-2}{2}}(w_{z}^{2})^{i}+\frac{%
w_{z}^{n}}{z},
\end{equation}%
using that $\sum_{i=0}^{m}a^{i}=\frac{1-a^{m+1}}{1-a},$ and that $%
s_{z}=[1-w_{z}]/z,$ it follows%
\begin{equation}
c_{n}(z)=\left( \frac{1-w_{z}}{z}\right) \frac{(1-w_{z}^{n})}{(1+w_{z})}+%
\frac{w_{z}^{n}}{z}.
\end{equation}%
The previous expression is valid for arbitrary $w_{z}.$ In the proposed
dynamics, $w_{z}=\gamma /(z+\gamma ),$ leading to%
\begin{equation}
c_{n}(z)=\frac{1}{z+2\gamma }+\frac{1}{z}\left( \frac{2\gamma }{z+2\gamma }%
\right) \left( \frac{\gamma }{z+\gamma }\right) ^{n}.
\end{equation}%
In the time domain this expression recovers Eq. (\ref{Recoherence}). From $%
\lim_{z\rightarrow 0}zc_{n}(z)=1,$ it follows $\lim_{t\rightarrow \infty
}c_{n}(t)=1.$ On the other hand, a Markovian limit is recovered as $%
\lim_{n\rightarrow \infty }c_{n}(z)=1/(z+2\gamma ).$


\begin{thebibliography}{99}
\bibitem{demo} The demonstration can straightforwardly be performed by
taking an arbitrary Lindblad rate equation \cite{LindbladRate} for a set of
auxiliary states $\{\rho _{R}(t)\}$ and demanding that $\mathrm{Tr}[\rho
_{R}(t)]=P_{R}(t)$ $\forall R,$ where the set of probabilities $\{P_{R}(t)\}$
obey the classical master equation~(\ref{MasterClassicalGen}).
\bibitem{initials} Taking arbitrary initial conditions in Eq. (\ref%
{LindbladRate}), $\rho _{k}(0)=\rho _{0}q_{k},$ with $\sum_{k=0}^{2}q_{k}=1,$
it follows $\rho _{0}(t)=q_{0}e^{-\gamma t}\rho _{0},$ $\rho
_{1}(t)=q_{1}e^{-\gamma t}\rho _{0}+q_{0}\gamma te^{-\gamma t}\sigma
_{z}\rho _{0}\sigma _{z},$ and $\rho _{2}(t)=\{q_{2}+q_{0}[1-e^{-\gamma
t}(1+\gamma t)]\}\rho _{0}+q_{1}(1-e^{-\gamma t})\sigma _{z}\rho _{0}\sigma
_{z}\rho _{0}.$ From Eq. (\ref{rates}), the rate $\gamma (t)$ in Eq. (\ref%
{LindbladDispersivo}) becomes $\gamma (t)=2\gamma \lbrack q_{0}(1-\gamma
t)+q_{1}]/[e^{\gamma t}(q_{0}-q_{1}+q_{2})-2q_{0}\gamma t+2q_{1}],$\ while $%
\lim {}_{t\rightarrow \infty }\int_{0}^{t}\gamma (t^{\prime })dt^{\prime
}=\ln [1/(q_{0}-q_{1}+q_{2})].$
\end{thebibliography}
\end{document}